\documentclass[psfig,useAMS,usenatbib]{mn2e}
\usepackage{aas_macros}
\usepackage{psfig}
\usepackage{graphicx}
\usepackage{multirow}

\title{The COSMOS Density Field: A Reconstruction Using Both Weak Lensing and Galaxy Distributions}

\author[A. Amara et al]
{A. Amara$^1$\footnotemark[1], S. Lilly$^1$,  K. Kova\v{c}$^1$, J. Rhodes$^{2}$, R. Massey$^{3}$,  
G.~Zamorani$^{4}$, 
C.~M.~Carollo$^1$, \newauthor
T.~Contini$^{5,6}$, 
J.-P.~Kneib$^7$,
O.~Le Fevre$^7$, 
V.~Mainieri$^8$, 
A.~Renzini$^9$,
M.~Scodeggio$^{10}$,\newauthor
S.~Bardelli$^{4}$,
M.~Bolzonella$^{4}$, 
A.~Bongiorno$^{11}$, 
K.~Caputi$^{1,21}$, 
O.~Cucciati$^{12}$,
S.~de la Torre$^{3}$, \newauthor
L.~de Ravel$^{3}$, 
P.~Franzetti$^{10}$,
B.~Garilli$^{10}$,
A.~Iovino$^{12}$, 
P.~Kampczyk$^1$,
C.~Knobel$^1$, \newauthor
F.~Lamareille$^{5,6}$, 
J.-F.~Le Borgne$^{5,6}$,
V.~Le Brun$^7$,
C.~Maier$^{1,20}$, 
M.~Mignoli$^{4}$, 
R.~Pello$^{5,6}$,\newauthor
Y.~Peng$^1$,
E.~Perez Montero$^{5,6,13}$,
V.~Presotto$^{12}$,
J.~Silverman$^{14}$, 
M.~Tanaka$^{14}$,
L.~Tasca$^7$,\newauthor
L.~Tresse$^7$,
D.~Vergani$^{4}$,
E.~Zucca$^{4}$,
L.~Barnes$^1$, 
R.~Bordoloi$^1$, 
A.~Cappi$^{4}$,
A.~Cimatti$^{15}$,\newauthor
G.~Coppa$^{11}$,
A.~Koekoemoer$^{16}$, 
C.~L\'opez-Sanjuan$^{7}$, 
H.~J.~McCracken$^{17}$, 
M.~Moresco$^{15}$,\newauthor
P.~Nair$^{4}$,
L.~Pozzetti$^{4}$, 
N.~Welikala$^{19}$\\
$^1$Institute for Astronomy, ETH Zurich, Zurich 8093, Switzerland,
$^{2}$Jet Propulsion Laboratory, California Institute of Technology, \\4800 Oak Grove Drive, Pasadena, CA 91109.
$^{3}$University of Edinburgh, Royal Observatory, Blackford Hill, Edinburgh, EH9 3HJ, UK.\\
$^{4}$INAF Osservatorio Astronomico di Bologna, via Ranzani 1, I-40127, Bologna, Italy,
$^{5}$Institut de Recherche en Astrophysique et Plan\'etologie, \\CNRS, 14, avenue Edouard Belin, F-31400 Toulouse, France,
$^{6}$IRAP, Universit\'e de Toulouse, UPS-OMP, Toulouse, France,
$^{7}$Laboratoire \\d'Astrophysique de Marseille, CNRS/Aix-Marseille Universit\'e, 38 rue Fr\'ed\'eric Joliot-Curie, 13388, Marseille cedex 13, France,
$^{8}$European \\Southern Observatory, Garching, Germany,
$^{9}$INAF - Osservatorio Astronomico di Padova, Vicolo dell'Osservatorio 5, 35122, Padova, Italy,\\
$^{10}$INAF - IASF Milano, Milano, Italy,
$^{11}$Max Planck Institut f\"ur Extraterrestrische Physik, Garching, Germany,
$^{12}$INAF Osservatorio \\Astronomico di Brera, Milan, Italy,
$^{13}$Instituto de Astrofisica de Andalucia, CSIC, Apartado de correos 3004, 18080 Granada, Spain,
$^{14}$Institute \\for the Physics and Mathematics of the Universe (IPMU), University of Tokyo, Kashiwanoha 5-1-5, Kashiwa, Chiba 277-8568, Japan,\\
$^{15}$Dipartimento di Astronomia, Universit\`a degli Studi di Bologna, Bologna, Italy,
$^{16}$Space Telescope Science Institute, Baltimore, Maryland 21218, \\USA,
$^{17}$Institut d'Astrophysique de Paris, UMR7095 CNRS, Universit\'e Pierre \& Marie Curie, 75014 Paris, France,
$^{18}$Max Planck Institut \\f\"ur Astrophysik, Garching, Germany,
$^{19}$Insitut d'Astrophysique Spatiale,  B\^atiment 121, Universit\'e Paris-Sud XI  \& CNRS, 91405 Orsay Cedex, \\France,
$^{20}$University of Vienna, Department of Astronomy, Tuerkenschanzstrasse 17, 1180 Vienna, Austria,
$^{21}$Kapteyn Astronomical Institute, \\University of Groningen, P.O. Box 800, 9700 AV Groningen, The Netherlands. 
}

\date{}

\begin{document}

\maketitle

\label{firstpage}

\begin{abstract}

\noindent The COSMOS field has been the subject of a wide range of observations, with a number of studies focusing on reconstructing the 3D dark matter density field. Typically, these studies have focused on one given method or tracer. In this paper, we reconstruct the distribution of mass in the COSMOS field out to a redshift $z=1$ by combining Hubble Space Telescope weak lensing measurements with zCOSMOS spectroscopic measurements of galaxy clustering. The distribution of galaxies traces the distribution of mass with high resolution (particularly in redshift, which is not possible with lensing), and the lensing data empirically calibrates the mass normalisation (bypassing the need for theoretical models). Two steps are needed to convert a galaxy survey into a density field. The first step is to create a smooth field from the galaxy positions, which is a point field. We investigate four possible methods for this: (i) Gaussian smoothing, (ii) convolution with truncated isothermal sphere, (iii) fifth nearest neighbour smoothing and (iv) a muliti-scale entropy method. The second step is to rescale this density field using a bias prescription. We calculate the optimal bias scaling for each method by comparing predictions from the smoothed density field with the measured weak lensing data, on a galaxy-by-galaxy basis. In general, we find scale-independent bias for all the smoothing schemes, to a precision of $10\%$. For the nearest neighbour smoothing case, we find the bias to be $2.51\pm 0.25$. We also find evidence for a strongly evolving bias, increasing by a factor of $\sim3.5$ between redshifts $0<z<0.8$. We believe this strong evolution can be explained by the fact that we use a flux limited sample to build the density field.

\end{abstract}


\section{Introduction}

The COSMOS field \citep{2007ApJS..172...38S} is the largest region of the sky that has been mapped contiguously with the Hubble Space Telescope (HST). This field has since been the subject of a wide range of studies aimed at measuring the detailed properties of objects in this field in many ways, including spectroscopic follow-up with the VLT \citep{2007ApJS..172...70L,2009ApJS..184..218L}, infrared imaging with Spitzer \citep{2007ApJS..172...86S}, x-ray observations with XMM and Chandara \citep{2007AAS...211.9532E,2007ApJS..172...29H}, UV imaging with GALEX \citep{2007ApJS..172..468Z}, Subaru imaging \citep{2007ApJS..172....9T} and radio observations with the VLA \citep{2007ASPC..375..123S}. The wealth of data coming from this region makes the COSMOS field perfect for developing techniques that bring together different data sets. These probe combination methods are useful for current studies, since it allows us to maximise the information coming from current data, as well as helping us to prepare for future wide-field surveys that will increasingly rely on probe combination \citep{2006astro.ph..9591A,2006ewg3.rept.....P} to make high precision measurements. For instance such surveys will aim to measure the dark energy properties, such as the equation of state, at the percent level.

We will focus in this paper on methods for reconstructing the density field from weak lensing and the spatial distribution of galaxies. This type of density field reconstruction work is a very active field and has been performed on a number of surveys. For instance, \cite{2009MNRAS.400..183K} performed a density reconstruction of the Sloan Digital Sky Survey (SDSS) north cap using data release 6 \citep{2008ApJS..175..297A} using Wiener Filtering. A number of density field reconstruction studies have also been performed on the COSMOS field. These include reconstructions of the density field using the galaxy distribution \citep{2010ApJ...708..505K,2011ApJ...731..102K} and weak lensing \citep{2007Natur.445..286M}. Each of these approaches has its own strengths and drawbacks. For instance, the density fields constructed from the weak lensing data alone have very poor resolution in redshift. This is due to a convolution along the line of sight by `the lensing efficiency function', which we discuss in more detail in section \ref{sec:WL}. The strength of the lensing maps, however, is that the weak lensing signal is a direct probe of the underlying matter. Maps constructed using galaxy positions rely on the galaxies acting as tracers of the matter field (see section \ref{sec:GB}). On the positive side, these galaxy position reconstructions have substantially better resolution in redshift than what is possible with weak lensing. Nonetheless, since the galaxies that we see are, at best, biased tracers of the underlying density field, additional assumptions and simplifications are needed to produce a reconstruction of the density field. 

The work presented here, therefore, aims to perform a matter reconstruction using both the galaxy position and weak lensing data. Such a combination allows us to construct a density field with high resolution in redshift that is calibrated from the data. We do this by measuring the expected lensing signal for each galaxy in the weak lensing survey that would come from a particular density field reconstruction. With this we can then look for correlations between the measured shear signals and predictions on a galaxy by galaxy basis. This approach to the problem is very powerful because it gives us freedom to choose how we average the data to reduce noise. For instance, we do not need to bin the data spatially to average out intrinsic shape noise. Instead we can average over predicted shear, which is both easier and more stable, or we can do away with binning entirely through direct correlations.

This paper is organised as follows. In section \ref{sec:overview}, we present an overview of some of the main issues associated with density reconstruction for galaxy clustering and weak lensing. We then present the COSMOS data sets in section \ref{sec:data} with a discussion of our methodology in section \ref{sec:method}. We give our results and conclusions in sections \ref{sec:res} and \ref{sec:conc}.

\section{Overview of the Issues}
\label{sec:overview}
When studying the density field, $\rho$, we typically focus our attention, in cosmology, on the over-density, $\delta$. For a given cosmic time $t$ this is deÞned as
\begin{equation}
\label{eq:den}
\delta = \frac{\rho - \bar{\rho}}{\bar{\rho}},
\end{equation}
where $\bar{\rho}$ is the mean cosmic density at that time. We do this because it is these perturbations of the smooth background density that drive structure formation and cause the bending of light rays.

\subsection{Galaxy Bias}
\label{sec:GB}

It is well known that since galaxies form at peaks in the background density field, the statistics of their distribution follow that of the underlying dark matter in a biased way. For instance, the two-point correlation of the peaks is boosted relative to the two-point correlation function of the underlying field, where the boost factor depends on the threshold used to define a peak. This boosting factor in the correlation function, which we will call Kaiser bias ($b_k$), was first identified by \cite{1984ApJ...284L...9K}, but it has since been calculated by a number of other works \citep{1988MNRAS.235..715E,1989MNRAS.237.1127C,1987MNRAS.227....1K,1985MNRAS.217..805P,1986ApJ...304...15B} for Gaussian random fields, as well as being studied using process named peak-background splitting (\cite{2010MNRAS.402..589M}. Since, for instance, in the Press-Schechter framework \citep{1974ApJ...187..425P}, the peaks above a critical threshold become halos and galaxies, the bias is often used to refer to the factor that links the galaxy and dark matter correlation functions,
\begin{equation}
\xi_g(r) = b_k^2(r)\xi_m(r),
\label{eq:kaiser}
\end{equation} 
where $\xi_g$ and $\xi_m $ are the two-point correlation functions of the galaxies and matter, respectively, as a function of separation distance $r$ at a given redshift. In principle, the Kaiser bias, $b_k$, could be scale dependent. With the bias defined in this way, its meaning and interpretation are straightforward and unambiguous. Both correlations can be measured in numerical simulations (assuming that the halos host the galaxies). This has been done by many authors \citep[for examples, see][]{2004ApJ...601....1W,1996MNRAS.282.1096M}. The general findings are that on large scales the bias is constant and results are consistent with expectations from the Halo model \citep{1996MNRAS.282..347M,2000ApJ...543..503M}, which links the halo mass to density threshold of the matter fields.

The link between the galaxy population and the underlying density field in real space is more complex due to the fact that galaxies form a set of points and not a smooth field. We then need to translate a set of coordinates in space (for each galaxy we have redshift $z$ and the two coordinates $\theta$ and $\phi$) into a real space galaxy density measure ($\delta^g$), which can then be used to build the smooth underlying matter density field ($\delta^m$). The steps, therefore are, 
\begin{equation}
\{\theta_i, \phi_i, z_i\} \rightarrow \delta^g(\theta, \phi, z) \rightarrow \delta^m(\theta, \phi, z).
\end{equation}
The relationship between the smoothed galaxy density field and the underlying matter field will then depend on the way that the smooth galaxy field was created. This means that for real space density reconstruction there will not be a unique bias. Instead the bias ($b_X$), where we use $X$ to denote the smoothing method, will depend on the particular scheme that has been chosen for going from galaxy positions to a continuous density distribution. For instance, producing a continuous galaxy field by smoothing with a Gaussian of fixed angular size, i.e. convolving a Gaussian and a set of delta functions, will likely require a different scaling ($b_{\rm Gauss}$) than a scheme with a dynamic smoothing scale, such as estimating the density using a fixed number of nearest neighbours ($b_{\rm NN}$). 

The simplest link between the continuous galaxy density field and the matter over-density is linear bias, where 
\begin{equation}
\label{eq:bias}
\delta^g_{X} = b_{X} \delta^m.
\end{equation}
This simple relation is similar to the Kaiser bias in equation (\ref{eq:kaiser}), except that the small scales will be affected by the smoothing used to go from galaxy points to the smooth density field. It is also possible to invoke more complex relationships, such as a non-linear and redshift dependent bias,
\begin{equation}
\label{eq:nonlinbias}
\delta^g_{X} = b_1(z) \delta^m + b_2(z) (\delta^m)^2,
\end{equation}
and in general the bias can be expected to be redshift dependent. Note that although we have omitted the subscript $X$, which would cause our notation to become clumsy, the biases are still specific to a given smoothing scheme.  This will be true for all real space biases (and inverse biases) used in this paper even if we omit the $X$ subscript for convenience.

The nonlinear bias in equation \ref{eq:nonlinbias} would likely lead to a scale dependent Kaiser bias. In all cases, the most accurate bias in real space will depend on (i) the tracers used, (ii) the smoothing scheme for going between galaxies as points and a smooth density field and (iii) the way that the density field will be used. On the last point, it is not clear that there exists a single density field that is optimal for all users. A density field to be used for cosmology may have different requirements than one used for galaxy evolution studies. For instance, for the former it may be best to use a simple filter so that the statistics of the matter field can easily be compared with prediction from theory, whereas the complex mapping from point to smooth field, e.g. involving dynamic smoothing scales, may be more optimal for the latter. 

\subsection{Weak Lensing}
\label{sec:WL}

Gravitational lensing effects are caused by the bending of light by intervening matter as it travels from source to observer. Because of this, lensing observables are integrated quantities along the line of sight. It is then difficult to make accurate measurements in the redshift direction. The advantage of weak lensing, however, is that the signal does not depend on the type of matter along the light's trajectory. This means that dark matter, which is the dominant  matter component, dominates the lensing signal. For this reason, lensing reconstructions of the matter density have focused mainly on 2D projected reconstructions \citep{1993ApJ...404..441K,2007A&A...462..459G,2008MNRAS.385..695B,2009ApJ...702..980K,2008MNRAS.385.1431H,2011arXiv1102.5743K}. Although the redshift dependence is weak, there have also been attempts to perform full 3D reconstructions using lensing data \citep{2002PhRvD..66f3506H,2003MNRAS.344.1307B,2007Natur.445..286M,2009MNRAS.399...48S}. These methods are based on weak lensing tomography, which employs the sensitivity of weak lensing signal to the source redshift.

In weak lensing, the most commonly used observable is cosmic shear. This is where the bending of the light rays causes the images of background galaxies to become sheared.  For instance, if a background galaxy is initially a circle, cosmic shear would distort the image into an ellipse. The effect is very subtle (causing typically percent level changes in the axis ratios of galaxies), so it must be studied statistically over very large samples of galaxies. Here, we give a very brief summary of the lensing basics and point the reader to the following review articles for more details: \cite{2003ARA&A..41..645R,2008PhR...462...67M,2001PhR...340..291B}.

The lensing induced effect that warps a galaxy image can be described to first order using the distortion matrix $A$
\begin{equation}
A =  (1-\kappa)\left(\begin{array}{cc}1-g_1 & -g_2 \\-g_2 & 1+g_1\end{array}\right),
\end{equation}
where $\kappa$ is known as the convergence, since the first term causes a change in the size of the image, and $g_1$ and $g_2$ are the two components of what is called the reduced cosmic shear, which causes an anisotropic shearing of the galaxy image. The reduced shear is related to the shear $\gamma$ through the expression
\begin{equation}
g = \frac{\gamma}{1 - \kappa}.
\end{equation}
Often in weak lensing we make the approximation that $\gamma = g$. However, it is worth noting that the actual observable is always reduced shear. The method we present here is able to properly account for the reduced shear. We find that for the accuracy possible with the COSMOS survey, approximating the reduced shear with the shear works well, but the distinction is likely to become more important for future surveys. 

Both shear and convergence can be related to each other through their dependence on the lensing potential $\psi$,
\begin{equation}
\label{eq:lenspot}
\kappa = (\psi_{,11} + \psi_{,22})/2 ;~\gamma_1 =  (\psi_{,11} - \psi_{,22})/2;~ \gamma_2 = \psi_{,12}, 
\end{equation}
where the subscripts denote a second order partial derivative such that $\psi_{,ij} = \partial^2\psi/\partial\theta_i\partial\theta_j$. Finally, in the weak limit typical of cosmic shear studies, where the light path is mildly perturbed, the convergence can be linked directly to the mass through an integration along the line of sight,
\begin{equation}
\label{eq:kappa}
\kappa = \frac{3H_0^2\Omega_m}{2c^2} \int_0^{\chi_s}  \frac{\chi (\chi_0 - \chi)}{\chi_0} \frac{\delta}{a(\chi)} d\chi,
\end{equation}
where $\chi$ is the comoving radial distance and $\chi_s$ is the distance to the source. This relies on a simplification known as the Born approximation. We can see from this expression that the strength of the lensing signal from a given over-density depends on the redshift of the background source that is being lensed.  In this way, the lensing signal can be seen as a radial convolution of the density with the lensing efficiency function.  Methods have been put forward for recovering this radial information and effectively performing a deconvolution (see \cite{2009MNRAS.399...48S} for further discussion). However, as with any deconvolution of noisy data, a perfect recovery of the original density field is not possible and the weak lensing reconstructed maps have very low resolution in redshift.

\subsection{Joint Analysis}

There are a number of ways that information coming from weak lensing data and the spatial distributions of galaxies can be combined. Two examples that are worth outlining are ones based on comparing statistical properties, such as correlation functions computed for galaxies and for the shear, and what is known as galaxy-galaxy lensing, which depends on the cross-correlation between the position of a foreground galaxy and the lensing of a background galaxy \citep{1996ApJ...466..623B,2001astro.ph..8013M,2004AJ....127.2544S,2006MNRAS.372..758M,2007arXiv0709.1159J,2009ApJ...703.2217S,2010ApJ...709...97L}.  The former approach is most useful when studying the galaxy biasing using a Kaiser-like bias (equation \ref{eq:kaiser}). This results naturally by measuring the correlation function of the galaxy distribution and comparing it to the correlation of the lensing, both of which can be measured directly. It is also useful in such an approach to compare different types of correlation functions, such as the aperture mass statistic \citep{2002ApJ...577..604H,1998ApJ...498...43S}. This is the approach taken by \cite{2012arXiv1202.6491J} in studying the bias properties of galaxies in the COSMOS field. 

The second approach of using galaxy-galaxy lensing has also been widely studied. An example of galaxy-galaxy lensing in COSMOS is presented by \cite{2011arXiv1104.0928L}. The strength of such a study is its ability to measure the average properties, such as the radial profile, of a given set of galaxies. However, the lensing effect of known secondary lenses along the line of sight typically is not included in the data analysis, instead it is included statistically as a `2-HALO term'. 

Our aim here is to create a real space density reconstruction. We take a more direct approach, outlined below in section \ref{sec:method}, to relate and calibrate the density of galaxies and the overall matter density field in real space. Our method, therefore, sits between the pure correlation function approach and the classic galaxy-galaxy lensing studies.

\section{COMOS/\lowercase{z}COMOS data}
\label{sec:data}
\subsection{The COSMOS Field}

Much of the data from the COSMOS field that we use here has already been presented and described in detail in the literature. Therefore, in the sections that follow we highlight some of the key features of this data, and we refer the reader to the appropriate sources for more of the technical details. 

\subsection{Weak Lensing Data: Hubble ACS Imaging} 

Our weak lensing analysis relies on detailed  measurements of the shapes of the galaxies in the COSMOS field. For this we use the shape catalogs generated from the HST ACS images. The catalogs are described in detail in \cite{2007ApJS..172..219L} and 
\cite{2007ApJS..172..203R}, with the pipeline having received a number of improvements since then. These updates are described in  \cite{2011arXiv1104.0928L} and include improvements in the treatment of Charge Transfer Inefficiency (CTI)  \citep{2010MNRAS.401..371M}. The lensing catalog is constructed from 575 ACS/WFC tiles with a total of $1.2 \times 10^6$ galaxies to a limiting magnitude of $I_{F814W}$ = 26.5. After lensing selection cuts are made, the final COSMOS weak lensing catalog contains $3.9 \times 10^5$ galaxies. These have accurate shape measurements and correspond to a density of 66 galaxies per square arc minute over the 1.64 square degree area covered by the lensing catalog. Figure \ref{fig:z_dist} shows the redshift distribution of our lensed galaxies along with curves showing the lensing efficiency function, see equations \ref{eq:kappa} and \ref{eq:kappa2}, for each of the samples.

\subsection{Weak Lensing Data: Photometric Redshifts}

The redshifts of the lensed galaxy sample have been measured through their photometry. For the work presented here, we use v1.8 \citep[updated from][]{2009ApJ...690.1236I}. These redshifts were determined using the 30 band multiwavelength data analysis presented in \cite{2009ApJ...690.1236I}. This included  deep $Ks$, $J$, and $u$ band data, which allows for accurate photo-z measurements at $z > 1$ through the 4000A break. More details on the data and the photometry can be found in \cite{2007ApJS..172...99C}.  The photo-z  measurements used a template fitting method (Le Phare)  that was calibrated with large spectroscopic samples from VLT-VIMOS and Keck-DEIMOS \citep{2007ApJS..172...70L,2009ApJ...690.1236I}. The dispersion in the photo-zÕs as measured by comparing to the spectroscopic redshifts is $\Delta z/(1+z _{spec}) = 0.007$ at $\rm i_{AB} < 22.5$, where $\Delta z = z_{spec} - z_{phot}.$ 

\subsection{Spectroscopic Redshifts}

For building the density field from the galaxy distribution, we rely on the zCOSMOS galaxy sample \citep{2007ApJS..172...70L,2009ApJS..184..218L}. We use galaxies with IAB $<$ 22.5 in the 1.1 deg$2$ region of the COSMOS field. Roughly $60\%$ of these galaxies have spectroscopically confirmed redshifts from the 20k zCOSMOS-bright sample, with the redshifts of the remaining galaxies having been determined photometrically and have typical errors of roughly $\delta z = 0.01(1+z)$. This provides us with a flux limited population. We have decided to use a flux limited sample because this has the advantage of containing the largest possible number of galaxies. An alternative option is to use a volume limited sample, which would likely have a simpler bias prescription. If the bias is introduced through a theory prior, this would be very useful. However, since our primary objective is to reconstruct the density field and we are able to measure the bias empirically using lensing, we are free to use any galaxy sample even though the bias may be complex. Since our measurement of the bias is a means to an end, it does not need to be easy to interpret. 

\begin{figure}
\begin{center}
\includegraphics[width=85mm]{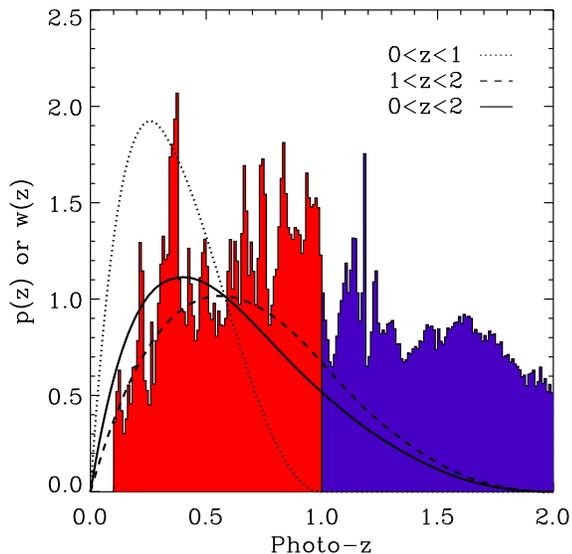}
\caption{The redshift distribution of the background lensed galaxies. The red sample shows the galaxies with photometric redshift of less than one and the blue sample are the galaxies with photometric redshifts between one and  two. The dotted, dashed and solid curves show the lensing efficiency function for the red, blue and both samples, respectively. Each of these curves has been normalised so that the area is one. The means of each curve give an effective redshift that is probed by that sample. These are $z_{eff} = 0.36$ (red sample), $z_{eff} = 0.70$ (blue sample) and$z_{eff} = 0.62$ (all galaxies).}
\label{fig:z_dist}
\end{center}
\end{figure}

\section{Our Methodology}
\label{sec:method}
\subsection{Building a 3D Density and Lensing Cube}
\label{sec:3Dcube}
Our objective in this study is to build the matter density field out to a redshift of $z=1$. This is similar to what was done in \cite{2010ApJ...708..505K}. We focus on the central region of the COSMOS field ($149.575 < R.A. < 150.675$ deg and $1.75 < DEC < 2.7$ deg), which is covered by both the weak lensing ACS catalog and the zCOSMOS data. We divide this volume into a 3D grid with a resolution of $256 \times 256 \times 500$. We then place galaxies into this volume based on their angular position and redshift. In the redshift direction we account for uncertainties in the redshift estimation by using the full probability distribution function (PDF) coming from the redshift estimation code. For the zCOSMOS galaxies, the redshift PDF is effectively a delta function, so these galaxies are assigned to a single pixel, while the photo-z galaxies are distributed over several pixels in redshift (depending on the photo-z error of a given galaxy). With this 3D galaxy density field, we can move towards calculating the convergence and shear inside this volume. To do this, we first need to convert the 3D galaxy density field ($\rho^g$) into a galaxy over-density field ($\delta^g$ - see equation \ref{eq:den}) \citep{2011ApJ...731..102K}. 

We recall, from equation \ref{eq:kappa}, that the convergence is given by the integral of the matter over density ($\delta^m$),
\begin{equation}
\kappa^p = \int \delta^m(z) w(z)  dz,
\label{eq:kappa2}
\end{equation}
where $\kappa^p$ would be our predicted convergence and $w(z)$ contains all the weight functions from equation \ref{eq:kappa}. This can be converted into a predicted 3D shear $\gamma^p$ field through the relationships shown in equation \ref{eq:lenspot}. This is similar to methods used in \cite{2009A&A...505..969P}. To link this calculation with our galaxy over density field ($\delta^g$), we need only to introduce the relationship between this and the matter over density ($\delta^m$), which we do through the bias (as discussed in section \ref{sec:GB}). Since the lensing observable comes from the matter over-density, it is simpler for our applications to reverse the usual bias relationship (equation \ref{eq:bias}). Instead we will use $\mu \equiv 1/b$ and work with
\begin{equation}
\delta^m = \mu \delta^g
\end{equation}
for constant bias and 
\begin{equation}
\delta^m(z) = \mu(z) \delta^g(z)
\end{equation}
for the analog of redshift dependent bias. Once again, note that we have dropped the subscript $X$ to simplify our notation. It is also possible to work with an analog of non-linear bias (equation \ref{eq:nonlinbias}), $\delta^m = \mu_1(z) \delta^g +  \mu_2(z) (\delta^g)^2$. However, for the work presented here we focus exclusively on linear bias. Clearly, in the simplest case of linear biasing, then $b = 1/\mu$. Here, we focus on reconstructions to first order (i.e. linear and only including $\delta^g$ terms). From this, we see that predicted convergence is thus given by
\begin{equation}
\kappa^p = \int \mu(z)\delta^g(z) w(z)  dz,
\end{equation}
and for the case where $\mu$ is independent of redshift 
\begin{equation}
\kappa^p = \mu \int \delta^g(z) w(z)  dz = \mu \kappa^g,
\end{equation}
where $\kappa^g$ is the convergence of the raw galaxy density field without a bias. We find that a parameterised expansion of $\mu$  in terms of $z/(1+z)$, such that $\mu(z) = \mu_{0} + \mu_{1}z/(1+z)$, gives a convenient form to explore redshift evolution. In this case, the convergence can be separated as 
\begin{equation}
\kappa^p= \mu_{0}\kappa^g +\mu_{1}\kappa^{\prime g},
\label{eq:k_evol}
\end{equation}
where $\kappa^{\prime g}$ is defined as 
\begin{equation}
\kappa^{\prime g} = \int \frac{z}{1+z} \delta^g(z) w(z)  dz. 
\end{equation}
In this separable way, we can easily investigate the different contributions from a constant bias term and test whether an additional evolving term improves the agreement with the lensing data. Also, it is useful to remember at this point that a given convergence field can be converted into a corresponding shear field using the relation shown in equation \ref{eq:lenspot}. In this case, $\kappa^p$ leads to $\gamma^p$, $\kappa^g \rightarrow\gamma^g$ and $\kappa^{\prime g}\rightarrow \gamma^{\prime g}$.

\subsection{Background Cosmology}

	To calculate the expected lensing signal from a given density field, we need to assume a background cosmology. Specifically, we need to assume a background expansion that relates redshift and radial distances. The rationale that we have chosen to adopt is to separate cosmology constraints coming from geometry and structure growth and to only use data coming from geometry measures to set our fiducial cosmology. We do this because constructing a density field primarily focuses on growth, so we try to limit the cross-talk between external data and internal measurements coming from COSMOS. We have chosen our fiducial cosmology based on the analysis by the WMAP team (available on the NASA WMAP website\footnote{http://lambda.gsfc.nasa.gov/product/map/current/ params/lcdm\_sz\_lens\_wmap7\_bao\_snsalt.cfm}), which combines (i) the seven year WMAP Cosmic Microwave Background (CMB) data \citep{2011ApJS..192...18K}, (ii) the Type Ia supernovae compilation from the extended SDSS dataset \citep{2009ApJS..185...32K} and (iii) Baryonic Acoustic Oscillations (BAO) \citep{2010MNRAS.401.2148P}.
Our fiducial cosmology parameters are $\Omega_m = 0.278$, $\Omega_\Lambda = 0.722$ and $h = 0.699$. The rest of the cosmology parameters, such as the equation of state $w$, are taken to be consistent with standard $\Lambda$CDM (e.g. $w=-1$).

\subsection{Investigation of the Impact of Smoothing}
\label{sec:smoothing}
The procedure that we outlined in section \ref{sec:3Dcube} for converting the galaxies positions into a 3D grid can be seen as a form of smoothing. However, this is a very minimal level of smoothing. We can see this when we consider that the number of pixels in our grid is $256\times256\times500 \approx 3.3\times10^{7}$, which is close to three orders of magnitude larger than the number of galaxies that we place in the grid. At this level the grid is extremely sparse (mostly zero) and it is very difficult to construct convergence and shear fields \citep[for some discussions of importance of smoothing in calculating lensing properties see][]{2006MNRAS.367.1367A,2007MNRAS.376..113A,2012MNRAS.419.3414M}. Further smoothing is therefore needed. 

We have explored four procedures for converting the galaxy point field into a smoothed continuous field. Each of these methods is a perfectly valid method for constructing a smooth density field from a point distribution of galaxies. In this paper we do not address the question of which method is best since this is likely to strongly depend on the reason for wanting to create a density field. Instead we focus on how lensing can be used to calibrate the density field once a given smoothing scheme has been chosen. Since each different smoothing scheme will produce a different realisation of the smooth density field $\delta_X$, where $X$ would denote the smoothing scheme, it is possible that different smoothing methods would have different optimal bias scaling ($b_X$). These smoothing methods we use are

\begin{itemize}

\item {\bf Gaussian Smoothing:} In this approach, the 2D mass field is convolved with a 2D Gaussian smoothing kernel. The results we show use a Gaussian with a standard deviation of 1.2 arcmin, which corresponds to 5 pixels in the tangential direction. 

\item {\bf Truncated Isothermal Sphere:} A natural extension to the generic Gaussian smoothing is to modify the smoothing kernel using the profile of a Truncated Singular Isothermal Sphere (TSIS). This kernel is given by  $\Sigma(\theta) = \sigma^2/\theta \tan^{-1}(\sqrt{\theta^2_T - \theta^2}/\theta)$, where $\Sigma$ is the 2D mass, $\sigma$ is the velocity dispersion for a Singular Isothermal Sphere (SIS) and is set by the mass of the TSIS. Again, for this study, all TSIS's have been normalised to have an area of one. Finally, we have set the truncation radius,  $\theta_T$, to 4.7 arcmin (20 pixels).

\item {\bf Nearest-Neighbor:} This is the approach adopted by \cite{2010ApJ...708..505K}. This method works by calculating the density of a given point based on the distance to the fifth nearest neighbors. In this way, the smoothing scale is adaptive and smooths over large scales in low density regions and small length scales in high density regions. \cite{2010ApJ...708..505K} have also implemented schemes for edge correction and treatment of mask, which they discuss in detail in their paper. Unlike the other three smoothing schemes that we have explored, this NN does not use the gridded galaxy distribution discussed in section \ref{sec:3Dcube}. Instead we use exactly the same procedure as in \cite{2010ApJ...708..505K}, which we remap to a $256\times256\times500$ grid so as to match the over-densities from our other methods.  

\item {\bf Multiscale Entropy Filter:} For a further alternative filtering scheme we also investigate the impact of using a multiscale entropy filter (MEF). This method has been implemented in the software package MRLens \citep{2006A&A...451.1139S}, which is publicly available. For those familiar with this method, we note that we decompose a given 2D mass sheet into eight wavelet scales and remove the first two scales (the first scale corresponds to the pixel scale). We note here that we have not attempted to modify the MEF implementation to optimize it for the noise properties of our kappa maps. Instead, we use the routines as they are from the MRLens package. This is likely to mean that our MEF smoothing is suboptimal. We, therefore, use the MEF filter as a point of comparison, but we will not be able to make a general statement about the absolute merit of MEF-like methods compared to our other smoothing schemes.

\end{itemize}

In principle, each of these filters could be applied to either  the 2D slices of the density field or the convergence field at each redshift, the latter being an integral quantity of the former. Since the aim of this present work is to construct the density field, we have focused on smoothing the 2D mass sheets.

\subsection{Using Predicted Shears}

By implementing the above approach we are able to calculate the predicted cosmic shear from a given density field at the position of every galaxy in the lensing catalog. For each of these galaxies, we would then have measured shear $\gamma_1$ and $\gamma_2$, noting that shear has two components and that the expected shear coming from the density field for each galaxy is $\gamma^p_1$ and $\gamma^p_2$. Let us assume then that 
\begin{equation}
\gamma_i = \gamma_i^p + \gamma_i^N,
\end{equation}
where $\gamma^N$ is a term that contains the noise contributions to the measured shear and $i$ can take the value of 1 or 2 for the two shear components. The predicted shear is linked to the shear coming from the galaxy density field ($\gamma_i^{g}$) through the inverse bias, $\gamma_i^p = \mu\gamma_i^{g}$. Finally, we can extend the calculation of the expected lensing signal out to galaxies at redshifts higher than $z=1$, i.e. outside our density data cube. These galaxies will be subjected to an additional lensing from structures between $z=1$ and the galaxy. However, if this structure is not correlated to the structure inside our cube, then this extra lensing signal is effectively an additional source of noise. This will likely be subdominant to other sources of noise, hence we ignore it.

\subsubsection{Fits to Predicted Shear}
\label{sec:fit}
The average measured shear should go to zero for large number of galaxies (i.e. $\langle\gamma_i\rangle \rightarrow 0$). However, if we bin the measured shear according to the predicted shear from the density field, then $\langle \gamma\rangle_{\gamma_p} \rightarrow  \gamma_p$. Therefore, for a good density reconstruction, a plot of $\langle \gamma\rangle_{\gamma_p}$ vs. $\gamma_p$ should, for the right bias, give a straight line with a gradient of one. In terms of $\chi^2$, we can use the quality of the fit as
\begin{equation}
\chi^2 = \sum \frac{(\langle \gamma\rangle_{\gamma_p} - \gamma_p)^2}{\langle \gamma^2\rangle_{\gamma_p}},
\end{equation}
where $\langle \gamma^2\rangle_{\gamma_p}$ is the variance of the data for fixed $\gamma_p$ bin, which will be dominated by the errors. This simple view is clear to understand if indeed the true density field is well described by the constant linear bias model in equation \ref{eq:bias}. In the case of more complex relations it is important to make the distinction between variations in the underlying density that are not correlated with the density field, which will be washed away in the averaging process, and deviations that are correlated with the galaxy density. An example of the latter would be a non-linear bias, such as the second term on the right-hand side of equation \ref{eq:nonlinbias}. For this reason, a more precise way to understand our procedure is that we are measuring the best-fit bias, using a linear approximation, that links the galaxy over-density field to the matter over-density field.  

\subsubsection{Zero-lag Covariance for Constant Bias}
\label{sec:zerolagcov}
As well as the simple fitting procedure outlined in the preceding section, it is also possible to make the bias measurements without needing to bin the data. We will show in the results section that both methods give consistent results. However, the advantage of removing the binning step is that calculations become more stable. This becomes especially important in the case of evolving bias, which we will discuss in the next section. The `no binning' approach is to look at the covariances between the measured and predicted data. Since we have two components  and two measures of shear, each galaxy gives a four element data vector $\{\gamma^g_1,\gamma^g_2,\gamma_1,\gamma_2\}$. We manipulate the elements of the data vector such that 
\begin{equation}
\tilde{\gamma}_i = \frac{\gamma_i}{\sqrt{\langle(\gamma_i^g)^2\rangle}} ;~ \tilde{\gamma}_i^g =\frac{\gamma_i^g}{\sqrt{\langle(\gamma_i^g)^2\rangle}}, 
\end{equation}
where $\sqrt{\langle(\gamma_i^g)^2\rangle}$ is the standard deviation of each of the shear components over all galaxies, to create a new data vector for each galaxy $\{\tilde{\gamma}^g_1,\tilde{\gamma}^g_2,\tilde{\gamma}_1,\tilde{\gamma}_2\}$.  For constant $\mu$, the (zero-lag) covariance matrix of these elements over all galaxies is

\begin{eqnarray}
\left(\begin{array}{cccc}
\langle\tilde{\gamma}_1^g\tilde{\gamma}_1^g \rangle & \langle\tilde{\gamma}_2^g\tilde{\gamma}_1^g\rangle  & \langle\tilde{\gamma}_1\tilde{\gamma}_1^g \rangle  & \langle\tilde{\gamma}_2\tilde{\gamma}_1^g \rangle  \\
- & \langle\tilde{\gamma}_2^g\tilde{\gamma}_2^g \rangle  & \langle\tilde{\gamma}_1\tilde{\gamma}_2^g \rangle  & \langle\tilde{\gamma}_2\tilde{\gamma}_2^g \rangle  \\
- & - & \langle\tilde{\gamma}_1\tilde{\gamma}_1 \rangle  & \langle\tilde{\gamma}_2\tilde{\gamma}_1 \rangle  
\\- & - & - & \langle\tilde{\gamma}_2\tilde{\gamma}_2 \rangle \end{array}\right) \\ 
=\left(\begin{array}{cccc}1 & 0 & \mu & 0 \\- & 1 & 0 & \mu \\- & - & 1+ \langle\tilde{\gamma}_1^N\tilde{\gamma}_1^N \rangle & 0 \\- & - & -  &  1+ \langle\tilde{\gamma}_2^N\tilde{\gamma}_2^N \rangle \end{array}\right) + N,
\label{eq:zerolagcov}
\end{eqnarray}
where the matrix shows the leading order terms and $N$ is a noise matrix that tends to zero as the total number of galaxies is increased. The covariance matrix, therefore, contains a wealth of information that we can use to measure $\mu$ and estimate its errors. 

\subsubsection{Zero-lag Covariance for Evolving Bias}

For the case with an evolving galaxy bias, we can analyze the data in similar ways. When working with the data, we found that while the $\chi^2$ method gives stable results for the case of constant bias, the results become unstable when the redshift bias is allowed to evolve. We find that this comes from the binning step and is due to the fact that the underlying signal is much weaker than the noise per galaxy. For the constant bias case, the relative ranking of the predicted shears of different lensed galaxies is maintained as we vary the bias. Therefore, when we bin in predicted shear, a given bin will contain the same lensed galaxies. This makes the calculations stable when varying the bias, but it does mean that we need to be cautious about the overall normalization of the $\chi^2$ functions. In the case of the evolving bias, the rank order of the predicted shears changes, which can change the compositions of the bins and cause erratic results becausee of the large noise terms. Instead, we have found that an extension of the zero-lag covariance method gives stable results. We also performed simple Monte-Carlo realsations of the data to confirm this effect. 

We can extend the zero-lag covariance method to evolving bias case by noting that, from equation \ref{eq:k_evol}, the predicted shear can be decomposed into two parts,
\begin{equation}
\gamma_{i}^{p} = \mu_0 \gamma_{i}^{g} + \mu_1 \gamma_{i}^{\prime g}.
\end{equation}
Here $\gamma_{i}^{p}$ are the two components of the predicted shears, $\mu_0$ and $\mu_1$ are the parameter expansions of the inverse bias. The shear prediction $\gamma_{i}^{g}$ and $\gamma_{i}^{\prime g}$ can be calculated from the density field directly and thus can be compared directly to the measures shear $\gamma_{i}$. By measuring the covariances of these three quantities we can construct $\mu_0$ and $\mu_1$, which are related by
\begin{equation}
\mu_0 = \frac{\langle\gamma_{i}\gamma_{i}^{g}\rangle}{\langle\gamma_{i}^{g}\gamma_{i}^{g}\rangle} - \mu_1\frac{\langle\gamma_{i}^{\prime g}\gamma_{i}^{g}\rangle}{\langle\gamma_{i}^{g}\gamma_{i}^{g}\rangle}
\label{eq:mu0}
\end{equation}
and
\begin{equation}
\mu_1 = \frac{\langle\gamma_{i}\gamma_{i}^{\prime g}\rangle}{\langle\gamma_{i}^{\prime g}\gamma_{i}^{\prime g}\rangle} - \mu_0\frac{\langle\gamma_{i}^{\prime g}\gamma_{i}^{g}\rangle}{\langle\gamma_{i}^{\prime g}\gamma_{i}^{\prime g}\rangle}.
\label{eq:mu1}
\end{equation}
We see from equations \ref{eq:mu0} and \ref{eq:mu1} that we have two equations with two variables, so we can easily solve for $\mu_0$ and $\mu_1$. The other advantage of this approach is that the errors on $\mu_0$  and $\mu_1$ can directly be calculated from combinations of fourth and second order moments of the data. For a discussion of these techniques see Chapter 5 of \cite{1993stp..book.....L}.

\section{Results}
\label{sec:res}

\subsection{Constant Bias}
\subsubsection{Constant Bias with $z<1$}

\begin{figure}
\begin{center}
\includegraphics[width=85mm]{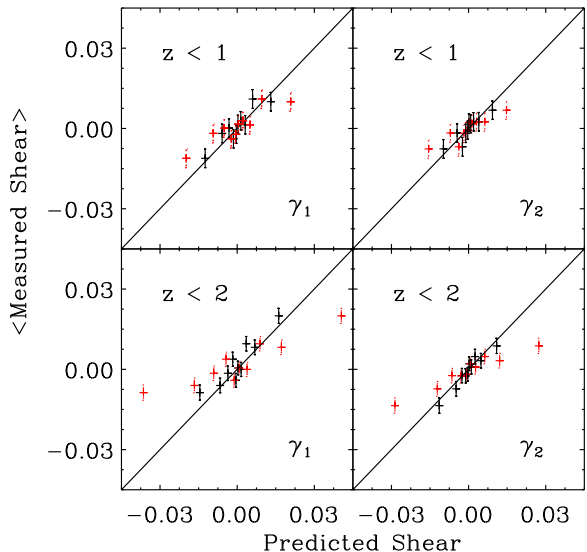}
\caption{The predicted shears vs measured shears for the Nearest-Neighbour (NN) smoothing cases. The top panels show results using lensed galaxies out to z=1 with a bias of 1.59, and the bottom panels show results that include lensed galaxies out to z=2, where we have set the bias to 2.51. The results for the two shear components are shown separately ($\gamma_1$ on the left and  $\gamma_2$ on the right). To build these results we bin the data in predicted shear. In the upper panel, each data point is an average of $\sim8000$ galaxies, and in the lower panel each data point is averaged over $\sim15000$ galaxies. The red results show what happens when a bias of one is used to make the shear predictions.}
\label{fig:shears}
\end{center}
\end{figure}

In our first test, we create a self consistent region out to redshift of $z=1$.  Inside this 3D volume we create the number density field, using the methods outlined in section \ref{sec:smoothing}, and the shear field by integrating the gradient of the lensing potential along the line of sight. By calculating the shear field inside this full volume, we are able to assign a shear to each of the galaxies in the COSMOS lensing catalog. The top panels of \ref{fig:shears} show a comparison between the predicted shear, which include the best fit bias, coming from the NN smoothed density field and the measured shear for lensed galaxies with redshifts less than one. In this figure, we have binned the data in predicted shear so that each data point contains roughly eight thousand galaxies. In this way, the random scatter coming from intrinsic shape noise is reduced. We see a clear correlation between the predicted shear and the measured shear once the best fit bias of 1.59 is used. The red symbols in background show the results when a bias of one is used. Note that the two shear components give separate measures of the lensing signal. It is therefore very important that  both give consistent results in figure \ref{fig:shears}. This is  equivalent to noting that the covariance matrix in equation \ref{eq:zerolagcov} contains two independent entries for $\mu$, one from $\gamma_1$ and another from $\gamma_2$, that should agree to within the random errors.

\begin{figure*}
\begin{center}
\includegraphics[trim=0.5cm 1cm 7cm 1.6cm, clip=true,angle = 90, width=160mm]{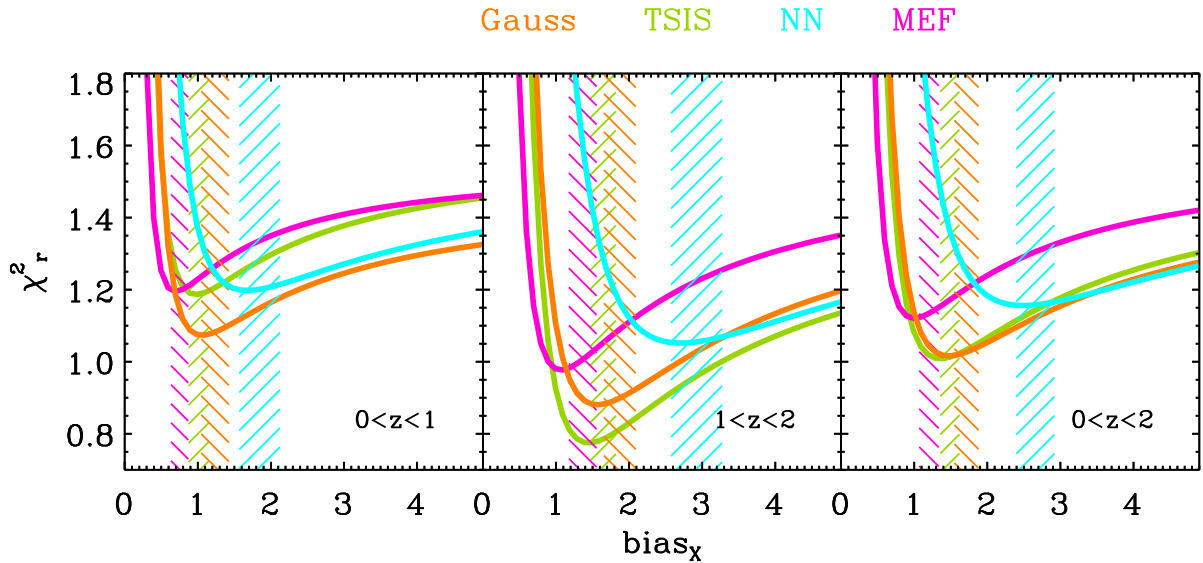}
\caption{The curves show the reduced $\chi^2$ fits for constant bias cases. As we have repeatedly highlighted, it is not a surprise that the best fitting bias depends on the specific scheme used to smooth the galaxy density field ($b_X$). The left panel is for galaxies in the range $0<z<1$. The middle panel is for galaxies in the range $1<z<2$, and on the right we see the results for all galaxies out to a redshift of $z=2$. The curves show the reduced $\chi^2$ results for each of our four smoothing cases: (i) Gaussian smoothing; (ii) convolution with a Truncated Singular Isothermal Sphere (TSIS); (iii) averaging of the fifth Nearest Neighbours (NN); and (iv) a Multiscale Entropy Filter (MEF). To calculate the reduced $\chi^2$, the lensed galaxies have first been divided into bins of width $\Delta z = 0.05$. Each bin has been further divided into 5 bins in predicted shear. The vertical shaded regions show the one sigma measurements coming from the zero-lag correlation method outlined in section \ref{sec:zerolagcov}. }
\label{fig:chi2z1}
\end{center}
\end{figure*}

\begin{figure}
\begin{center}
\includegraphics[trim=3.3cm 2cm 0.cm 1cm, clip=true,width=85mm]{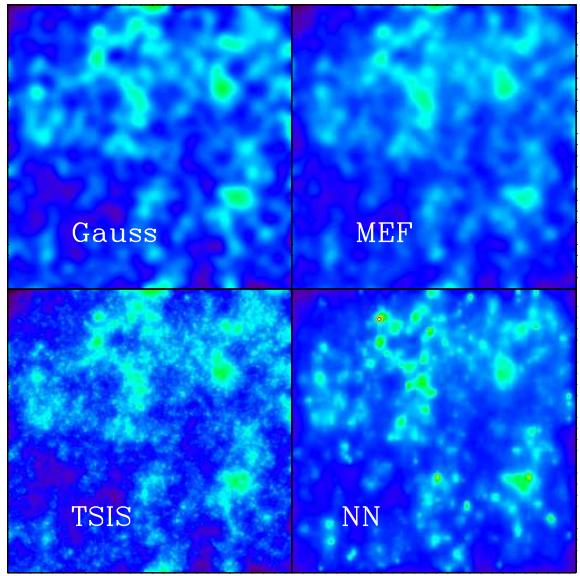}
\caption{The four panels show the convergence ($\kappa$) of the COSMOS field at a redshift of 1 for the four smoothing cases that we study (see Table \ref{tbl:res}). Each panel shows the central 1.1 square degrees of the COSMOS field. The colour range is the same for all panels and corresponds to -0.04 (blue) to 0.1(red).}
\label{fig:convergence}
\end{center}
\end{figure}

To find the best fit constant bias for this sample we perform both the $\chi^2$ and the zero-lag covariance methods discussed in sections  \ref{sec:fit} and \ref{sec:zerolagcov}. The curves in the left panel of figure \ref{fig:chi2z1} show the reduced $\chi^2$ as a function of bias for our four smoothing cases. It is worth noting that the same calculations for no smoothing does not give a good fit (or a minimum) and stays substantially out of the range of the plot for any value of the bias. We also show, in the background, shaded regions that come from the measurements using the cross-correlations method. We see that the two methods for measuring the bias give good agreement. The best fit bias does depend on the smoothing scheme that we adopt and that in each case we typically measure the bias to a precision of 10\% to 15\%, once a smoothing scheme has been chosen. Specific best-fit results and the one sigma errors for the zero-lag covariance method are shown in table \ref{tbl:res}. We also restate that for the lensed galaxies in the range $0 < z< 1$, the effective redshift, i.e. the mean of the lensing efficiency function shown in figure \ref{fig:z_dist}, is $z_{eff} = 0.36$. In figure \ref{fig:convergence} we show the convergence field at $z=1$ coming from our four reconstructed densities. We see that all the maps show the same broad features, but they vary in the details. Thanks to the peaked nature of the TSIS smoothing kernel, we see that this reconstruction has more small-scale features than the Gaussian and MEF maps. Though the nearest neighbor convergence field has a variance that is similar to the other maps, we can see from the bottom right panel that the positive extremes are larger. This can be attributed to the fact that the nearest neighbor scheme has a dynamic smoothing scale, and we are, therefore, able to resolve the highest density peaks, which contain many galaxies.

\begin{table}
  \centering 
\begin{tabular}{| l | c | c |  c | }
\hline
\multirow{3}{*}{Smoothing} & \multicolumn{3}{c}{Bias per slice:}  \\
& $z_{eff} = 0.36$&  $z_{eff} = 0.70$& $z_{eff} = 0.62$\\
& [ $0<z_s<1$ ]&  [ $1<z_s<2$ ]& [ $0<z_s<2$ ]\\
\hline
Gauss 	& 	$ 1.20\pm 0.19$	&   $ 1.85\pm 0.22$	&	$ 1.70\pm 0.17$ \\
TSIS 	&     	$ 0.99\pm 0.14$	&   $ 1.62\pm 0.18$ 	&	$ 1.47\pm 0.13$ \\
NN 		&  	$ 1.59\pm 0.22$  	&   $ 2.82\pm 0.34$ 	&	$ 2.51\pm 0.25$ \\
MEF 		&   	$ 0.73\pm 0.12$ 	&   $ 1.34\pm 0.19$ 	&	$ 1.19\pm 0.13$\\
\hline
\end{tabular}
\caption{Best fit values and errors for each of the four smoothing that we have investigated using the covariance method outlined in section \ref{sec:zerolagcov}. Results are shown for the same redshift slices as in figure \ref{fig:chi2z1}. These have been constructed by dividing the source (i.e. lensed) galaxies using the photometric redshift ($z_s$). For each of these redshift slices we also show the effective redshift ($z_{eff}$).}
\label{tbl:res}
\end{table}

\subsubsection{Constant Bias with $z<2$}

Although we restrict our density reconstructions to redshifts of less than one, we are still able to use the lensing data at higher redshifts. This is because the additional lensing signal from the mass in the redshift range $1 < z < 2$ will effectively behave as an additional noise term. Furthermore, this additional noise term will be subdominant to the noise coming from the random orientation of galaxies and  will not have a significant impact on our analysis. The only concern would be if there was a strong correlation of the mass distribution in these two regimes. However, given the distances involved ($\sim$ Gpc), this cross-correlation is likely to be subdominant. 

The middle panel of figure \ref{fig:chi2z1} shows results using the lensed galaxies in the redshift range $1<z<2$ and the right panel shows results for the redshift range $0<z<2$. Here again, the galaxies are divided into bins with widths $\Delta z = 0.05$, and in each redshift slice the galaxies are divided into 5 bins in predicted shear. The results are also summarised in table \ref{tbl:res}.  For each of the redshift intervals, we draw similar conclusions to those from the low redshift sample. First, there is good agreement between our two approaches for measuring the bias and the best-fit bias depends on the smoothing scheme. We also see that, by using the full sample, we are able to measure the bias to a precision of roughly 10\%, once a particular scheme is chosen for going between the galaxy field and continuous field. The precision for a given scheme seems to be at this level regardless of which one is considered. We also see that for all cases the measured bias increases with the effective redshift ($z_{eff}$) being probed. Since this trend is independent of our smoothing scheme, we can conclude that the bias of our underlying sample must increase with redshift. 

\subsection{Redshift Evolving Bias}

As we outlined in section \ref{sec:method}, it is computationally more simple to work with the inverse of the bias, $\mu(z)$, that is expanded in a series. Given the expansion shown in equation \ref{eq:k_evol}, the bias is then given by 
\begin{equation}
b(z) = \frac{1}{\mu_0 + \mu_1y},
\end{equation}
where $y = z/1+z$. When expressed in this form, we see that we must take care in how we explore possible values of $\mu_0$ and $\mu_1$, since there is a danger that the bias could become singular, for instance at $\mu_0 (1+z) =  - \mu_1 z$ or negative. To guard against these cases, it is convenient to recast our variable such that
\begin{equation}
b(z) = \frac{1}{\mu_0 (1 + f y)} = \frac{b_0}{1 + f y},
\end{equation}
where $f$ is given by $f = \mu_1 /\mu_0$. Sensible bounds can now be easily placed on $f$ to ensure that the bias is positive and non-singular.

\begin{figure}
\begin{center}
\includegraphics[width=85mm]{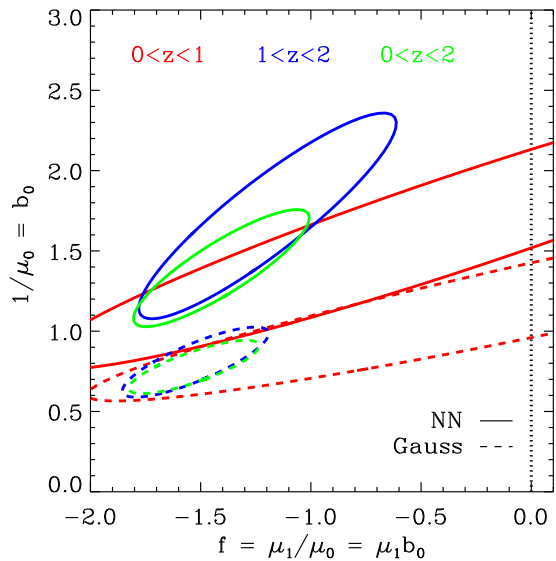}
\caption{The one sigma measurements for the bias normalization $b_0 = 1/\mu_0$ and the term $f$ that controls the redshift evolution. This factor is defined as $f = \mu_1/\mu_0$ and can also be expressed in terms of the constant bias $f = b_0 \mu_1$. A constant non-evolving bias would have $f=0$, which is shown with a dotted line. The results shown here are for the cases with fifth nearest neighbour (solid) and Gaussian (dashed) smoothing. We show the constraints to the low redshifts (red), high redshifts (blue) and the full sample out to a redshift of 2 (green).} 
\label{fig:res_evol}
\end{center}
\end{figure}

Figure \ref{fig:res_evol} shows the one sigma constraints on the two parameters $f$ and $b_0$ for the nearest neighbor (solid) and Gaussian (dashed) smoothing schemes. For each case, we show the measurements for the three redshift configurations shown in table \ref{tbl:res}. Once again, we see that the exact normalization of the bias depends on the smoothing scheme, but both cases show evidence for an evolving bias. We also see that when only using galaxies in the range $0 < z< 1$ we are able to measure an overall amplitude, but the constraints on the evolution of the bias are very weak. When we include the higher redshift sample, we see that the evidence for an evolving bias becomes very strong and that the constant bias option (dotted line at $f=0$) is clearly ruled out. 

\begin{figure}
\begin{center}
\includegraphics[width=85mm]{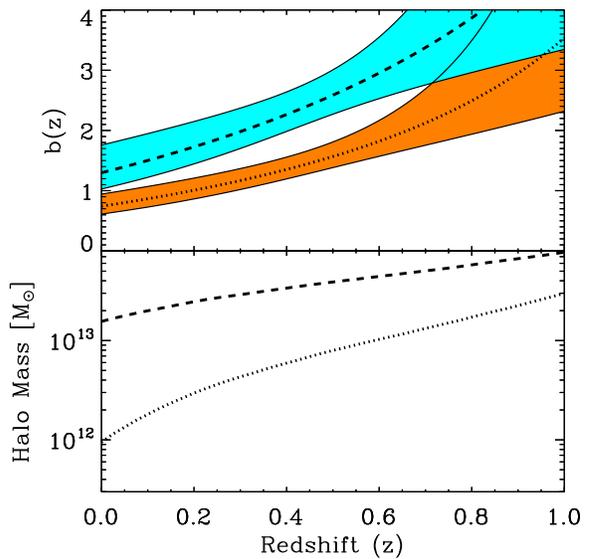}
\caption{The top panel shows the measured bias as a function of redshift coming from the one sigma errors in figure \ref{fig:res_evol}. Shown are the fifth nearest neighbour (cyan) and the Gaussian smoothing (orange) results. The bottom panel shows the mass threshold that would be necessary in a Mo \& White (1996) bias model to reproduce the best-fit biases shown in the upper panel.}
\label{fig:res_evol2}
\end{center}
\end{figure}

In the top panel of figure \ref{fig:res_evol2}, we show the one sigma bounds of the bias as a function of redshift ($b(z)$) for the NN and Gaussian smoothing. In each case we also show the best-fit biases. To gain an insight into the significance of the bias evolution, we explore the expected bias from a simple bias model. For this, we use the bias model presented by \cite{1996MNRAS.282..347M}. The bottom panel shows the mass limit in this model that would correspond to the best-fit biases shown in the upper panel. We must be careful in interpreting these curves, since the bias that we measure is averaged, using different smoothing prescriptions, in ways that are not accounted for in the analytic model. However, we can conclude that, to be consistent with expectations, there needs to be a strong evolution in the tracer with redshift. This is not unexpected for a flux limited sample. 

\section{Conclusions}
\label{sec:conc}
We have created a 3D density reconstruction of the COSMOS field out to a redshift of one using a combination of the galaxy clustering and weak lensing measurements. This empirical approach means that we do not need to make a priori assumptions about the relationship between galaxies and dark matter. Instead, we measure this directly while creating the optimal density field from galaxy tracers. The advantage of relying on a local tracer, such as the density of galaxies, rather than an integrated tracer, such as weak lensing, is that we are able to produce a density field that has a high resolution in redshift. 

We find that if we use only the lensed galaxies that are embedded in our density field, i.e. galaxies for which we are able to make a full prediction of their lensing signal,  then we are able to measure a constant non-evolving bias to a precision of 10\% to 15\%, once a smoothing scheme has been chosen for going from galaxy points to density field. However, we also argue that it is possible to use lensed galaxies outside of our density field since the extra lensing signal due to the mass at redshifts greater than one will have a very weak correlation to the mass in our density. For this reason, this extra lensing signal will act as an additional source of noise that will be sub-dominant to the noise from other sources. In this way, we are able to measure the bias to 10\% by using lensed galaxies out to a redshift of 2. This is encouraging as we look forward to future surveys with areas much greater than that covered by COSMOS. A good rule of thumb is that parameter errors typically scale as the inverse square root of the survey area. In this case, the next generation of surveys that will have areas over several hundred to thousands of square degrees can expect to reach sub-percent level precision using this method.

We have looked for evidence of an evolving bias. We find that for our density field, which is constructed using a flux limited sample, does have a strongly evolving bias. This trend remains irrespective of the smoothing scheme that we use to go from galaxy points to a smooth density distribution.

The work presented here is our first implementation of a method that draws a sharp distinction between local tracers and integrated measures. In our case, the local tracer is the number density of galaxies, which we use to construct the matter density field, and the integrated measure is the weak lensing signal, which we use to calibrate the free parameters in the reconstruction. It is then, in principle, easy to extend this method to include further local tracers, such as x-ray flux, that can be used to mold the density field and extra integrated quantities, such as strong lensing and SZ measurements, that can be used to test the accuracy of the reconstruction.

\section*{Acknowledgements}
AA would like to thank Alexandre Refregier and Julien Carron for useful and insightful conversations. This work was performed in part at JPL, run by Caltech for NASA. We would also like to thank the anonymous referee for his/her careful and timely reading of the first draft of the paper.

\bibliographystyle{mn2e}
\bibliography{/Users/amaraa/Work/Mypapers/mybib}

\end{document}